\documentclass[aps,pra,twocolumn,groupedaddress,showpacs]{revtex4}
\usepackage{amsmath,amssymb}
\usepackage{graphicx}
\begin{document}

\title{Diamagnetic nature of stratified thin metals in visible range}

\author{Masanobu Iwanaga}
\email{M.Iwanaga@osa.org}
\affiliation{Department of Physics, Graduate School of Science, Tohoku 
University, Sendai 980-8578, Japan}

\date{\today}

\begin{abstract}
It is numerically demonstrated that effectively strong diamagnetic resonance 
emerges at visible frequencies in stratified metal-dielectric metamaterials. 
The effective optical constants are extracted by two-complex 
reflectivity method. It is clarified that 
the effective diamagnetic response originates from local diamagnetism at 
stratified thin metals. 
The effective diamagnetism is crucially sensitive to the sturucture 
of unitcell. 
The effective diamagnetic response is always associated 
with effective plasma frequency and is therefore regarded as a magnetic 
component of the collective excitation. 
\end{abstract}


\maketitle

\section{Introduction\label{intro}}
Photonic metamaterials are trials to invent and produce novel 
electromagnetic (EM) states by modifying periodic structures 
in which constituents have the same optical constants with the bulk 
materials. The trials started at gigahertz, and has 
proceeded to terahertz and optical frequencies. The rapid 
progress was reviewed in \cite{Pendry04,Shal07}. 
The strategy is significantly different from nano-scale materials science 
in which quantum size effect on electronic states is a key. 
In metamaterials, novel effective EM states are explored by controlling the 
geometrical structures. The effective EM states are solutions of Maxwell 
equations for a homogenized system obtained by coarse-graining and represent 
averaged EM fields approximately. 

Recent metamaterial optics usually employs a retrieval way to evaluate 
effective optical constants \cite{Smith02}. The procedure is briefly 
explained as follows: When one 
knows complex transmissivity and reflectivity (called $S$ parameters) for a 
finitely thick object, one can numerically calculate (or retrieves) the 
effective permittivity and permeability by substituting the $S$ 
parameters into the transmission and reflection formulas, which hold for
a finitely thick material of effective optical constants. 
The way was first introduced in 2002 \cite{Smith02} and has been called 
retrieval way. 

The retrieval way is algorithmic and not rigorous compared with the effective 
media description, which has been studied over one century 
\cite{Lamb80,Berg80PRL,Fel05PRL} since Maxwell-Garnet \cite{MxG04,MxG06}. 
The effective media descriptions are usually derived for specified structures 
by assuming analytical models and the long wavelength limit. 
The difference between the effective media description and recent retrieval 
way was already described in the introduction of \cite{Smith05PRE}. 
In short, the effective media description is derived analytically for 
ideally large objects under the long wavelength approximation, 
whereas the retrieval way transforms the $S$ parameters into 
effective optical constants in a practical way. 

In the retrieval way, there are two distinct points from the effective media 
description: 
(i) the interface of metamaterials play a crucial role because the $S$ 
parameters are sensitive to the surface layers; 
(ii) the description using effective optical constants turned out to be 
approximately valid for several artificial metamaterials beyond the long
wavelength limit \cite{Smith05PRE}. 
By use of the retrieval way, effective magnetic responses and 
negative refraction at optical frequencies have been demonstrated 
experimentally \cite{Zhang05PRL,Shal05OL,Dol06OL}. 
Those metamaterials exhibit novel optical states, which are exceptions 
to the statement 
that permeability is unity at optical frequencies \cite{Landau}. 

A simple theoretical analysis presented a possibility of super-resolution 
by a thin metal film \cite{Pendry00PRL} and the 
sub-diffraction-limited image was shown experimentally \cite{Fang05Sci}. 
It is expected that near-EM-field enhancement is induced to compensate 
the loss at resonance with exploiting thin metallic layers. 
Another experimental result on transmission was reported which supports 
the enhancement of EM fields by stratified thin metals \cite{Scal98}. 
It was also shown that the enhancement is associated with high magnetic 
field in thin metals \cite{LZh05PRL}. 
Besides, it was recently argued that negative refractive response 
appears in stratified stacks composed of metal-dielectric several layers 
\cite{Scal07,Zhang07OE}. 
Thus, the magnetic response in metallic thin films attracts great interest, 
connecting to novel optical phenamena. 

\begin{figure}[t]
\begin{center}
\includegraphics[width=7.5cm,clip]{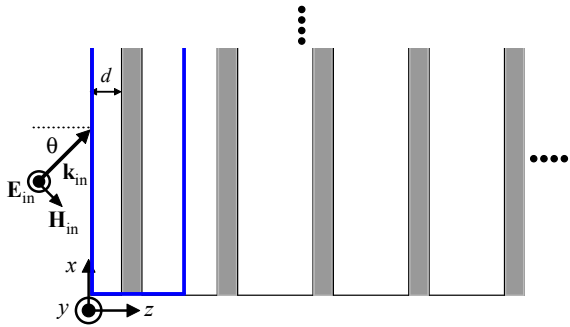}
\caption{Schematic drawing for structure of SMDM and coordinate configuration. 
Gray indicates metal (Ag) and white dielectric (MgF$_2$). 
Blue bold line indicates the unitcell of SMDM. 
Incident light sheds on $xy$ plane with $s$ polarization (\textit{i.e.} 
$\textbf{E}_{\rm in}||y$) and incident angle $\theta$. 
The thickness of surface layer is denoted by \textit{d}. 
In this configuration, the two products of $\varepsilon_y/\mu_x$ and 
$\mu_x \mu_z$ are determined from two incident angles. Details are given in 
Sec.\ \ref{TCRM}. \label{fig1}}
\end{center}
\end{figure}

In this paper, stratified metal-dielectric metamaterials 
(SMDM) are specified to clarify the effective magnetic response at visible 
frequencies. Typical optical properties of SMDM have been shown in 
an earlier publication \cite{Iwa07OL}. Here, the effective magnetic response 
is focused on and explored with modifying the structure of unitcell. 
Figure \ref{fig1} shows the schematic drawing of SMDM and the 
coordinate configuration. The metal is set to be Ag and the dielectric is 
MgF$_2$. The thickness of metal and dielectric are 15 and 60 nm, respectively. 
The periodicity is then 75 nm. The thickness of Ag 
layers was set to ensure using dielectric constant of bulk silver in 
literature \cite{Johnson}. 

In Sec.\ \ref{TCRM}, two-complex reflectivity method (TCRM) are described. 
Numerical results are shown in Sec.\ \ref{Result}, and the implication is 
explored to elucidate the origin of effective magnetic response. More 
discussion and comparison with other types of magnetism reported so far 
are given in Sec.\ \ref{discussion}. 

\section{Two-complex reflectivity method\label{TCRM}}
In this study, bulk SMDM is analyzed which is thick enough to eliminate 
transmission, and full components of $\varepsilon$ and $\mu$ tensors are 
extracted by TCRM. 
Although the scheme of TCRM was reported already 
\cite{Iwa07OL}, the full description is provided in the following. 

Since the structure of SMDM is uniaxial, effective tensors of 
permittivity $\varepsilon$ and permeability $\mu$ are 
assumed to be diagonal and uniaxial: 
\begin{equation}
\varepsilon = %
\begin{pmatrix}
\varepsilon_x & 0 & 0 \\
0 & \varepsilon_y & 0 \\
0 & 0 & \varepsilon_z 
\end{pmatrix},\,\,\,
\mu = %
\begin{pmatrix}
\mu_x & 0 & 0 \\
0 & \mu_y & 0 \\
0 & 0 & \mu_z 
\end{pmatrix}.\label{eps_mu}
\end{equation}
The optical axis is set to be the $z$ axis, and then the two relations of 
$\varepsilon_x = \varepsilon_y$ and $\mu_x = \mu_y$ hold 
in Eq.\ (\ref{eps_mu}). 

In linear and axial media, electric and magnetic flux densities 
are expressed using permittivity and permeability tensors as 
\begin{eqnarray}
\textbf{D} &=& \varepsilon_0\varepsilon\textbf{E}, \label{D=eE} \\
\textbf{B} &=& \mu_0\mu\textbf{H}, \label{B=mH} 
\end{eqnarray}
where $\varepsilon_0$ and $\mu_0$ are permittivity and permeability in 
vacuum, respectively. 

First, the boundary conditions of Maxwell equations are taken into 
account. In Fig.\ \ref{fig1}, incident wavenumver vector ${\bf k}_{\rm in}$ 
is in the $xz$ plane and then refracted wave has the wavevector 
${\bf k}$ in the $xz$ plane. We here introduce normalized refracted 
wavevector $\hat{\bf k}={\bf k}/k_0$ where $k_0$ is wavenumber 
of light in vacuum, and can write $\hat{\bf k}$ as 
\begin{equation}
\hat{\bf k} = %
\begin{pmatrix}
\hat{k}_x \\
0 \\
\hat{k}_z
\end{pmatrix}. \label{k-hat}
\end{equation}
Then, complex reflectivity $r_s$ under $s$ polarization (that is, 
$\textbf{E}_{\rm in}||y$) in Fig.\ \ref{fig1} is easily derived from the 
Maxwell boundary conditions \cite{Jackson}; the equation for $r_s$ is 
sometimes called Fresnel formula \cite{BornWolf}. 
After simple modification, the next equation is obtained: 
\begin{equation}
\frac{\hat{k}_z(\theta)}{\mu_x} = \frac{n_0\cos\theta}{\mu_0}%
\frac{1 - r_s(\theta)}{1 + r_s(\theta)}, \label{kz_s}
\end{equation}
where $\theta$ is incident angle, and $n_0$ denotes the refractive 
index in vacuum.  
In particular, $\hat{k}_z(0)$ is refractive index along the $z$ axis. 

Next, the equation of dispersion is included in TCRM. 
When the EM fields are monochromatic plane waves and 
proportional to $\exp(i\textbf{k}\cdot\textbf{r} - i\omega t)$, 
the two of the Maxwell equations are 
\begin{eqnarray}
i\textbf{k} \times \textbf{E} &=& i\omega\textbf{B}, \label{kxE} \\
i\textbf{k} \times \textbf{H} &=& -i\omega\textbf{D}. \label{kxH}
\end{eqnarray}
Note that the fields $\textbf{E}$, $\textbf{H}$, $\textbf{D}$, and 
$\textbf{B}$ in Eqs.\ (\ref{kxE}) 
and (\ref{kxH}) are independent of $(\textbf{r},t)$. 

When the solution of plane wave is assumed, it is generally allowed 
that the $\varepsilon$ and $\mu$ tensors depend on the frequency $\omega$ 
and wavevector ${\bf k}$; that is, the $\varepsilon$ and $\mu$ tensors 
have frequency and spatial dispersions 
($\varepsilon = \varepsilon(\omega,{\bf k})$ and 
$\mu = \mu(\omega,{\bf k})$) \cite{Landau}. 
However, it is for the present assumed that the $\varepsilon$ and $\mu$ 
tensors are independent of {\bf k}. In solid crystals, 
the $\mu$ tensor is set to be unity at optical frequencies and independent of 
{\bf k}, and the spatial dispersion of $\varepsilon$ tensor is often so weak 
that it is frequently negligible \cite{Landau}. 

Equations (\ref{D=eE}) and (\ref{kxH}) yield Eq.\ (\ref{pre_E_eq}), 
and Eq.\ (\ref{E_eq}) is derived with help of Eqs.\ 
(\ref{B=mH}) and (\ref{kxE}): 
\begin{eqnarray}
\varepsilon\textbf{E}%
 &=& -\frac{1}{\varepsilon_0\omega}%
\textbf{k} \times \textbf{H} \label{pre_E_eq} \\
 &=& -\frac{1}{\varepsilon_0\omega}%
\textbf{k} \times \left[ \frac{1}{\mu_0\omega}\,\mu^{-1}%
(\textbf{k} \times \textbf{E})\right] \label{E_eq} \\
 &=& -\hat{\textbf{k}} \times \left[ \mu^{-1}%
\left(\hat{\textbf{k}} \times \textbf{E}\right)\right]. \label{n_E_eq}
\end{eqnarray}
Equation (\ref{n_E_eq}) is obtained by using 
$\textbf{k} = k_0 \hat{\textbf{k}}$ and the relation 
\begin{equation}
\frac{k_0^2}{\varepsilon_0 \mu_0 \omega^2} = 1. \nonumber
\end{equation}
The determinant of coefficient matrix for \textbf{E} in 
Eq.\ (\ref{n_E_eq}) has to be zero to have nontrivial solution; therefore, 
Eq.\ (\ref{n_E_eq}) yields a general constraint between 
$\varepsilon$ and $\mu$ for a given $\hat{\textbf{k}}$. 
The constraint condition is equation of dispersion and 
can be regarded as generalization of Fresnel equation. Historically, 
Fresnel equation was derived under the condition of $\mu=1$, gave the 
equation concerning $\varepsilon$ for a given $\hat{\textbf{k}}$, 
and has been usually employed to explain optical phenomena, 
\textit{e.g.} birefringence, in solid crystals \cite{Landau}. 

Under \textit{s} polarization, Eq.\ (\ref{n_E_eq}) 
is modified using Eq.\ (\ref{eps_mu}) and (\ref{k-hat}) : 
\begin{eqnarray}
\varepsilon%
\begin{pmatrix}
0 \\
E_y \\
0
\end{pmatrix}
&=& 
-\hat{\textbf{k}}\times\left[ \mu^{-1}%
\begin{pmatrix}
-\hat{k}_z E_y \\
0 \\
\hat{k}_x E_y
\end{pmatrix}\right] \nonumber \\
&=& 
\begin{pmatrix}
0 \\
\mu_x^{-1} \hat{k}_z^2 E_y + \mu_z^{-1} \hat{k}_x^2 E_y \\
0
\end{pmatrix}. \nonumber 
\end{eqnarray}
We thus obtain the equation of dispersion under $s$ polarization, 
\begin{equation}
\varepsilon_y = \frac{\hat{k}_z(\theta)^2}{\mu_x} + %
\frac{\hat{k}_x(\theta)^2}{\mu_z}, \label{disper_s}
\end{equation}
where $\hat{k}_x(\theta) = n_0 \sin\theta$ in Fig.\ \ref{fig1}. 

Combining Eqs.\ (\ref{kz_s}) and (\ref{disper_s}), 
we obtain the next equation: 
\begin{equation}
\frac{\varepsilon_y}{\mu_x} = \left( %
\frac{n_0\cos\theta}{\mu_0}%
\frac{1 - r_s(\theta)}{1 + r_s(\theta)}\right)^2 + %
\frac{n_0^2 \sin^2\theta}{\mu_x \mu_z}. \label{master_eq}
\end{equation}
When one knows the value of complex reflectivity $r_s$ by optical measurement 
or computation, Eq.\ (\ref{master_eq}) is equation for 
$\varepsilon_y/\mu_x$ and $\mu_x \mu_z$. 
Two different angles ($\theta_1 \neq \theta_2$) yield 
$\sin\theta_1 \neq \sin\theta_2$, and enable to evaluate 
$\varepsilon_y/\mu_x$ and $\mu_x \mu_z$ uniquely. 

By permutating the configuration $(x,y,z) \to (y,z,x)$, we can figure 
$\varepsilon_z/\mu_y$ and $\mu_y \mu_x$. Then, incident light sheds on 
the $yz$ plane and has the polarization of ${\bf E}_{\rm in}||z$. 
The plane of incidence is the $xy$ plane. 
In this second step, Eq.\ (\ref{master_eq}) is modified as 
\begin{equation}
\frac{\varepsilon_z}{\mu_y} = \left( %
\frac{n_0\cos\phi}{\mu_0}%
\frac{1 - \tilde{r}_s(\phi)}{1 + \tilde{r}_s(\phi)}\right)^2 + %
\frac{n_0^2 \sin^2\phi}{\mu_y \mu_x}, \label{master_eq_2nd}
\end{equation}
where $\tilde{r}_s$ is complex reflectivity on the $yx$ plane and $\phi$ is 
incident angle in the $xy$ plane. 
In evaluating $\varepsilon_z/\mu_y$ and $\mu_y \mu_x$ 
similarly to $\varepsilon_y/\mu_x$ and $\mu_x \mu_z$, 
it is required that 
the object should be semi-infinitely thick along the $x$ axis. 

\begin{figure}[t]
\begin{center}
\includegraphics[width=7.5cm,clip]{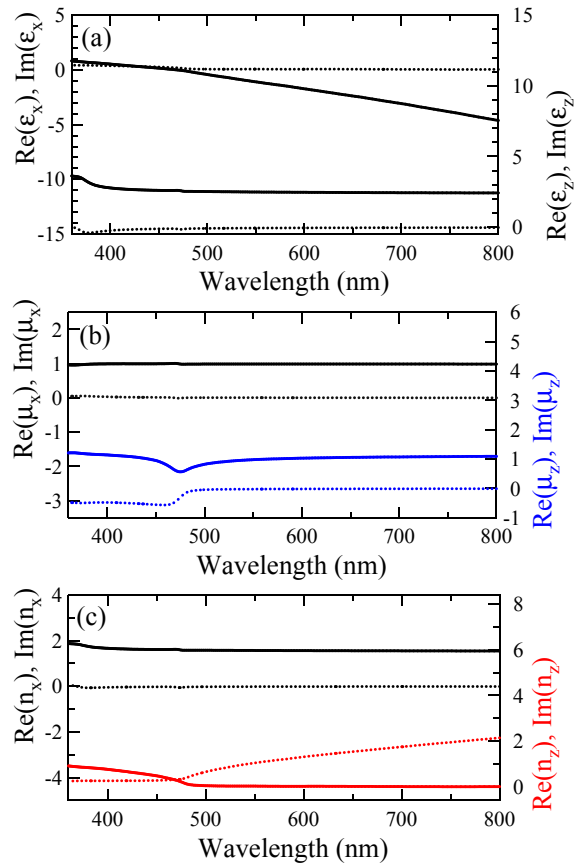}
\caption{Effective optical constants of SMDM, extracted with the two incident 
angles of 0$^{\circ}$ and 15$^{\circ}$. The surface thickness $d$ of 
SMDM is 20 nm. 
(a): Effective permittivity. Upper solid line denotes Re($\varepsilon_x$) 
and upper dotted line does Im($\varepsilon_x$). 
Lower two lines represent $\varepsilon_z$ likewise. 
(b): Effective permeability. Lines indicate $\mu_x$ and $\mu_z$ 
with the notation similar to (a).
(c): Effective refractive index along the $x$ and $z$ axes. The real parts 
are shown with solid lines and the imaginary parts with dotted lines. 
\label{fig2}}
\end{center}
\end{figure}

Since the relations of $\mu_x = \mu_y$ are assumed for SMDM, 
$\mu_x^2$ is evaluated in the second step; succeedingly, the rest 
components of $\varepsilon_x$, 
$\varepsilon_z$, and $\mu_z$ are uniquely obtained except for the 
sign. The sign of $\mu_x$ is set to be $\mu_x \approx 1$ at 800 nm because 
the $\mu_x$ spectrum keeps constant and 
does not show any resonant behavior around the wavelength. 
The sign of $\mu_x$ at different wavelength 
is taken to avoid discontinuous jump as shown in Fig.\ \ref{fig2}. 

Complex reflectivity $r_s$ was computed exactly, whereas $\tilde{r}_s$ was 
evaluated quite precisely, that is, within a-few-percent numerical error. 
The numerical way to compute $\tilde{r}_s$ was reported earlier 
\cite{Iwa07OL}. 
The precision in the computation for complex reflectivities ensures that 
the full components of tensors are determined within a-few-percent numerical 
error. 

\section{Numerical results\label{Result}}

\subsection{TCRM Analysis}
Figure \ref{fig2} presents the full components of (a) $\varepsilon$ 
and (b) $\mu$ tensors of SMDM of surface thickness $d = 20$ nm.
The tensors are extracted by TCRM with the two angles of 0$^{\circ}$ 
and 15$^{\circ}$. Other sets of two incident angles 
result in the same spectra 
of $\varepsilon_x$, $\varepsilon_z$, and $\mu_x$ within the numerical error. 
This consistency indicates that $\varepsilon_x$, $\varepsilon_z$, and $\mu_x$ 
are independent of incident angle and consequently wavevector {\bf k}. 

In TCRM analysis for SMDM using the angle pairs of ($0^{\circ}$,$\theta$) 
and ($0^{\circ}$,$\phi$), 
it is easily verified from Eq.\ (\ref{master_eq}) that $\mu_z$ depends only 
on the $\theta (\neq 0^{\circ})$. 
The three components of $\varepsilon_x$, $\varepsilon_z$, and $\mu_x$ could 
in principal depend on the incident angles; however, 
they have been found to be independent of incident angles and therefore 
to be well-defined for SMDM. On the other hand, 
the component $\mu_z(\theta)$ is exceptionally evaluated for each 
non-zero $\theta$. Thus, 
all the components of $\varepsilon$ and $\mu$ tensors are determined 
self-consistently and in a well-defined manner at visible frequencies 
in Fig.\ \ref{fig2}. 

The $\varepsilon_x$ component has typical dispersion of Drude metal 
(upper solid and dotted lines in Fig.\ \ref{fig2}(a): 
the real and imaginary parts of $\varepsilon_x$, respectively), 
and the effective plasma frequency $\omega_{\rm p,eff}$, which is defined 
by Re$[\varepsilon_x(\omega_{\rm p,eff})] = 0$, 
is located at $\lambda_{\rm p,eff} = c/2\pi\omega_{\rm p,eff} = 468$ nm. 
On the other hand, the $\varepsilon_z$ spectrum exhibits 
typical dielectric properties for the incident light of 
$\textbf{E}_{\rm in}||z$ on the $xz$ or $yz$ planes (lower solid and dotted 
lines in Fig.\ \ref{fig2}(a): the real and imaginary parts of 
$\varepsilon_z$, respectively). Concerning permeability components, 
$\mu_x$ is close to 1 at 360--800 nm (upper solid and dotted lines in 
Fig.\ \ref{fig2}(b): the real and imaginary parts of $\mu_x$, respectively), 
while $\mu_z$ goes down to 0.57 at 477 nm and shows strong diamagnetism 
(lower solid and dotted lines in Fig.\ \ref{fig2}(b): 
the real and imaginary parts of $\mu_z$, respectively). 
Refractive indexes along the $x$ and $z$ axes are displayed in Fig.\ 
\ref{fig2}(c); the upper solid and dotted lines respectively denote Re($n_x$) 
and Im($n_x$), and the lower solid and dotted lines respectively stand for 
Re($n_z$) and Im($n_z$). The imaginary parts of $n_x$ and $n_z$ are small 
below $\lambda_{\rm p,eff}$, and correspond to efficient EM-wave propagation 
in SMDM. 

\begin{figure}[t]
\begin{center}
\includegraphics[width=7.5cm,clip]{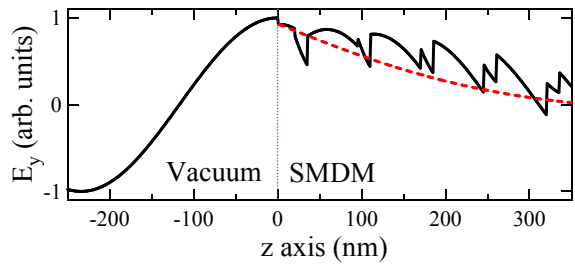}
\caption{Electric field vs the averaged field. Electric field is calculated 
at $\lambda_{\rm p,eff}$ under normal incidence on the $xy$ plane with $s$ 
polarization. Solid line indicates forward propagation only, that is, 
incident and refractive components. Dashed line represents the 
electric field calculated by the effective refractive index $n_z$ in Fig.\ 
\ref{fig2}(c). Vertical dotted line indicates the interface.  
\label{fig3}}
\end{center}
\end{figure}

In Fig.\ \ref{fig3}, we test the validity of description by effective 
refractive index at $\lambda_{\rm p,eff}$. The interface of vacuum and SMDM 
is set at $z=0$ and incident light illuminates on the $xy$ plane normally. 
Solid line denotes electric field of incident light in vacuum ($z < 0$); 
it represents refractive component of electric field in SMDM ($z > 0$). 
Dashed line presents the effective field profile calculated by using $n_z$ 
in Fig.\ \ref{fig2}(c) and reproduces the averaged field profile in SMDM 
approximately. Thus, TCRM extracts averaged EM response in SMDM. As was 
already confirmed in \cite{Iwa07OL}, the effective description has relevant 
physical meanings and works fairly well in the visible range of interest. 

\subsection{Origin of effective diamagnetism\label{origin}}
\begin{figure}[t]
\begin{center}
\includegraphics[width=8.0cm,clip]{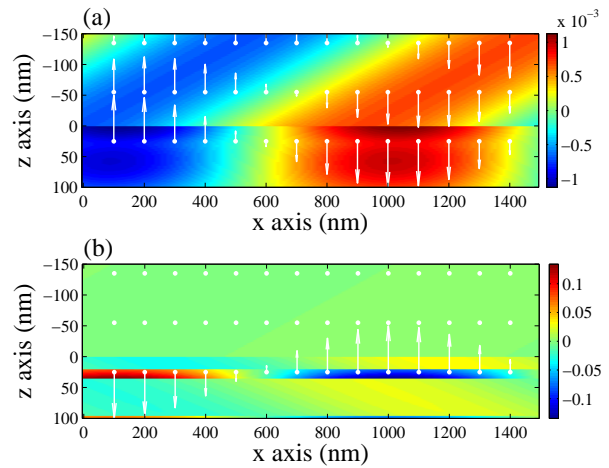}
\caption{Comparison of (a) $B_z$ and (b) $H_z$ at 477 nm, the minimum of 
Re($\mu_z$). The region of vacuum ($z<0$) shows incident light only, 
for simplicity; the incident angle is $\theta = 15^{\circ}$. 
The region of SMDM ($z>0$) contains 
both forwardly and backwardly propagating components. Arrows indicate 
the vectors at typical points (dots). In particular, the directions of $B_z$ 
and $H_z$ in metallic layer ($20<z<35$ nm) are clearly opposite. 
\label{fig4}}
\end{center}
\end{figure}

As shown in Fig.\ \ref{fig2}, effective diamagnetic response appears 
in the $\mu_z$ spectrum. 
This implies that magnetic induction $B_z$ and magnetic field $H_z$ behave 
in a different way. Figure \ref{fig4} displays the actual distribution of (a) $B_z$ and (b) $H_z$ as image plot. Arrows denote the vectors at dotted points. 
The wavelength is at 477 nm, where Re($\mu_z$) is minimun. 
In vacuum ($z<0$), the incident light travels with incident angle 
15$^{\circ}$; only incident component is shown for simplicity in Fig.\ 
\ref{fig4}. In SMDM 
($z>0$), both forward and backward components are displayed simultaneously 
because reflectance $R$ is not small ($R \sim 0.7$) 
and the backward components connected to inner multi-reflection 
is not negligible. 
The computation for $B_z$ at each point is carried out from Eq.\ (\ref{kxE}), 
and on the other hand magnetic field $H_z$ is calculated from 
Eq.\ (\ref{kxH}). 

Figure \ref{fig4} obviously shows that the directions 
of $B_z$ and $H_z$ are opposite in metallic layer ($20<z<35$ nm). 
Dielectric layers behave like paramagnetic materials. Besides, 
the $H_z$ component in the metallic layer is far more intense than 
the $H_z$ in vacuum and dielectric layers. In this way, the thin metallic 
layers are found to be diamagnetic along the $z$ axis. 
The diamagnetic nature does not come from electronic spin states of silver 
but from the superimposed EM fields peculiar to mesoscopic structure of SMDM. 
The backward component is indeed responsible for the local diamagnetism. 
Thus, the effective diamagnetic $\mu_z$ response is found to originate from 
the thin metallic layers. 

The diamagnetic response of metallic layer in SMDM is confirmed numerically 
in Fig.\ \ref{fig4}. It is also easily derived from Maxwell equation. 
In Fig.\ \ref{fig4}, Eqs.\ (\ref{kxE}) and (\ref{kxH}) in a metallic layer 
are explicitly written as 
\begin{eqnarray}
i\omega B_z &=& ik_0 \hat{k}_x E_y, \label{kxE_comp} \\
{\rm sgn}(\hat{k}_z)ik_0\hat{k}_z H_x - ik_0\hat{k}_x H_z &=& %
-i\omega\varepsilon_{\rm Ag}E_y, \label{kxH_comp}
\end{eqnarray}
where sgn($\hat{k}_z$) takes $1$ or $-1$, and denotes the forward and 
backward  propagation for the $z$ axis. 
From the computation for actual EM distribution, $|H_z|$ is found handreds 
times larger than $|H_x|$; \textit{i.e.} $|H_z|\gg|H_x|$. 
Consequently, the first term in the left hand side of Eq.\ (\ref{kxH_comp}) 
is negligible to a good approximation. Then, deviding Eq.\ (\ref{kxE_comp}) 
by Eq.\ (\ref{kxH_comp}), the sign of the real part of $B_z/H_z$ turns out 
to be determined by a factor $\varepsilon_{\rm Ag}$: 
\begin{equation}
{\rm Re}(B_z/H_z) \propto {\rm Re}(\varepsilon_{\rm Ag}).
\end{equation}
At the wavelength in Fig.\ \ref{fig4}, ${\rm Re}(\varepsilon_{\rm Ag}) < 0$ 
directly results in ${\rm Re}(B_z/H_z) < 0$ in the metallic layer. 
Thus, the diamagnetic nature of thin metallic layers is confirmed from the 
simple calculation. 

\subsection{Properties of $\mu_z$ resonance}
\begin{figure}[t]
\begin{center}
\includegraphics[width=7.5cm,clip]{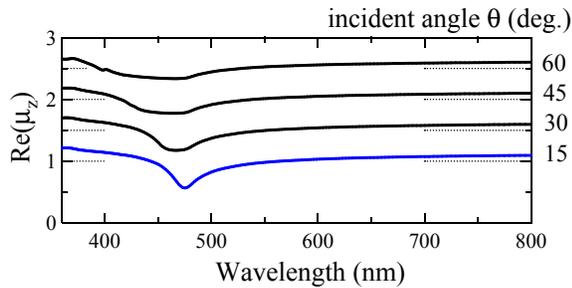}
\caption{Spatially dispersive behavior of $\mu_z$ resonance. 
All the spectra are obtained by TCRM with the two angles of 0$^{\circ}$ and 
$\theta(\neq 0^{\circ})$. However, as described in the text, $\mu_z$ depends 
only on the $\theta$ in Fig.\ \ref{fig1}. The spectrum at 15$^{\circ}$ is 
the same with that in Fig.\ \ref{fig2}(b). 
The spectra at $\theta = 30^{\circ}$, 45$^{\circ}$, and 60$^{\circ}$ are 
offset for visibility by 0.5, 1.0, and 1.5, respectively. Horizontal 
dotted lines indicate Re($\mu_z$) $=$ 1.0 for each spectrum. 
\label{fig5}}
\end{center}
\end{figure}

Figure \ref{fig5} shows the dependence of $\mu_z$ on 
incident angle. The $\mu_z$ spectra in Fig.\ \ref{fig5} vary for incident 
angles and indicates spatial dispersion. 
Note that $\mu_z$ at off-resonance above 540 nm is almost independent of 
$\theta$. The spatial dispersion is a resonant behavior. 

\begin{figure}[b]
\begin{center}
\includegraphics[width=7.5cm,clip]{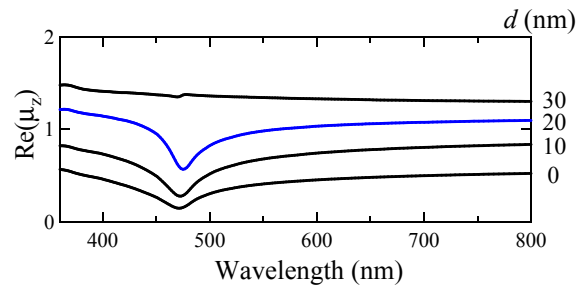}
\caption{Permeability $\mu_z$ spectra of SMDM with various unitcells: 
the thickness $d$ of surface layers from 0 to 30 nm. All the spectra 
are computed at the incident angle of 15$^{\circ}$, 
and are presented without vertical offset. The spectrum of 
$d = 20$ nm in Fig.\ \ref{fig5} is displayed again for comparison. 
\label{fig6}}
\end{center}
\end{figure}

Figure \ref{fig6} displays the effect of structure of unitcell on 
the effective 
$\mu_z$. Figures \ref{fig2}--\ref{fig5} are the results regarding the 
surface thickness $d=20$ nm; the spectra is shown again in Fig.\ \ref{fig6}. 
From bottom to top, the thickness $d$ varies from 0 to 30 nm, respectively. 
By changing the surface thickness, the structure of unitcell in SMDM varys; 
see Fig.\ \ref{fig1}. 
SMDM with $d=0$ and 10 nm show diamagnetic property in a wide 
range. The increase of surface thickness $d$ gradually reduces diamagnetism 
and finally eliminates the diamagnetism at $d=30$ nm. 
Since the structure of unitcell directly affects the phase of EM wave in SMDM, 
it is crucially significant how to set the unitcell in realizing the 
effective diamagnetism. SMDMs of different unitcells are different 
metamaterials. Figure \ref{fig6} is the explicit indication. It is also 
consistent with the conclusion in a recent numerical study \cite{Deco06PRL}, 
showing that the interfaces play a vital role in photonic slabs of finite 
thickness, in extracting effective constants by the retrieval way. 

As is shown in Figs.\ \ref{fig2} and \ref{fig6}, 
the effective diamagnetic response emerges at about $\lambda_{\rm p,eff}$, 
comes from the stratified metallic layers in SMDM. 
In order to observe the diamagnetic 
resonance in SMDM, it is essential how to select 
the structure of unitcell. 
As the $\lambda_{\rm p,eff}$ shifts with varying the ratio of metal in the 
unitcell, the diamagnetic resonance follows the shift of 
$\lambda_{\rm p,eff}$. It strongly suggests that the diamagnetic resonance is 
magnetic part of the effective plasma resonance. 
The spatially dispersive behavior in Fig.\ \ref{fig5} also shows a typical 
characteristic of resonance. 

\section{Discussion\label{discussion}}
From computational and theoretical studies, it was reported that effective 
magnetism 
takes place in the systems composed of dielectric \cite{Fel05PRL,Huang04} 
and metallic rods \cite{Fel05OL,Hu06}, results from the Mie resonances 
peculiar to the rods, and is spatially dispersive \cite{Fel05PRL,Fel05OL}. 
These magnetic responses stem from the local resonance due to geometrical 
shape of constituents. 

The effective diamagnetism in SMDM also inherited from the the local 
diamagnetic nature of 
thin metals, which was shown in Sec.\ \ref{origin}. There is 
a remarkable characteristic to the diamagnetism of SMDM, 
which is absent in the systems 
of the local Mie resonance. In the Mie systems it is known that 
the resonances always induce not a small amount of absorption loss, whereas 
the diamagnetic resonance is almost free from loss as seen in Fig.\ 
\ref{fig2}(c). Although small loss [{\it i.e.} small Im($n_z$)] exists, 
it appears irrespective of the diamagnetism; indeed, the small loss 
also exists in the non-magnetic SMDM of surface thickness $d=30$ nm in 
Fig.\ \ref{fig6} \cite{Iwa07OL}. 
This kind of magnetic resonance with low loss has not been found so far 
to our knowledge and is a new class of magnetism in metamaterials. 
Consequently, it is expected that low-loss, effective magnetic 
resonance at visible frequencies will be realized based on this magnetism. 

Thick enough SMDM have been focused on in this paper. Of course, 
this system is not easily fabricated in reality. 
However, it is a good touchstone to examine 
how the effective optical responses emerge in periodic metal-dielectric 
systems. In the present analysis, there is no evidence to suggest negative 
refraction in SMDM, which was lately argued in finite systems 
\cite{Scal07,Zhang07OE}. 
If negative refraction takes place in finitely thick SMDM, it would be 
due to an interference effect in finite systems. 

\section{Conclusions\label{conclusion}}
The full components of effective tensors of $\varepsilon$ 
and $\mu$ in SMDM has been numerically extracted by TCRM in a 
well-defined way. As a main result, the effective strong diamagnetic 
resonance along the optical axis has been clarified. 
The resonance is spatially dispersive. 
It has been revealed that stratified thin metals in SMDM are the source 
of effective diamagnetism. 
The diamagnetism of thin metals results not from local Mie resonances 
but from the EM states realized in SMDM. 
The diamagnetic resonance is induced always near the effective plasma 
frequency; it can therefore be regarded as a magnetic component of the 
collective excitation. 
The diamagnetic nature of thin metals is experimentally 
feasible, and it will be a good test to measure nonlinear optical responses 
via diamagnetic resonance. 

\begin{acknowledgments}
The author thanks T.\ Ishihara and S.\ G.\ Tikhodeev for stimulating 
discussion. 
This study was partially supported by Research Foundation for 
Opto-Science and Technology and by the Information Synergy Center, 
Tohoku University in numerical implementation.
\end{acknowledgments}

\end{document}